\documentstyle[12pt,epsfig]{article}

\newcommand{\beq}{\begin{equation}}
\newcommand{\eeq}{\end{equation}}
\newcommand{\bea}{\begin{eqnarray}}
\newcommand{\eea}{\end{eqnarray}}

\def\tev{\;{\rm TeV}}
\def\gev{\;{\rm GeV}}
\def\mev{\;{\rm MeV}}
\def\cN{{\cal N}}
\def\cM{{\cal M}}
\def\cO{{\cal O}}
\def\cF{{\cal F}}
\def\LQ{{\rm LQ}}

\def\half{{\textstyle{1 \over 2}}}
\def\lapprox{\lower .7ex\hbox{$\;\stackrel{\textstyle <}{\sim}\;$}}
\def\gapprox{\lower .7ex\hbox{$\;\stackrel{\textstyle >}{\sim}\;$}}

\begin{document}
\titlepage
\begin{flushright}
{DTP/97/32}\\
{hep-ph/9705229}\\
{May 1997}\\
\end{flushright}
\begin{center}
\vspace*{2cm}
{\Large {\bf Hadronic Antenna Patterns as a Probe of \\[2mm]
Leptoquark Production at HERA}} \\

\vspace*{1.5cm}
M.\ Heyssler$^a$ and W.\ J.\ Stirling$^{a,b}$ \\

\vspace*{0.5cm}
$^a \; $ {\it Department of Physics, University of Durham,
Durham, DH1 3LE }\\

$^b \; $ {\it Department of Mathematical Sciences, University of Durham,
Durham, DH1 3LE }
\end{center}

\vspace*{4cm}
\begin{abstract}
Hadronic antenna  patterns can provide a valuable  diagnostic tool for
probing the origin of the  apparent  excess of high $x,Q^2$  events at
HERA.  We present  quantitative  predictions for the  distributions of
soft particles and jets in standard deep inelastic  scattering $eq \to
eq$ events and in events  corresponding  to the production of a narrow
colour--triplet  scalar  (`leptoquark')  resonance.  The corresponding
distributions  of soft photon  radiation,  which are  sensitive to the
leptoquark  electric charge, are also presented.  The distribution for
one particular  leptoquark  assignment is shown to contain a radiation
zero.
\end{abstract}

\newpage

\section{Introduction}

The  observation of an apparent  excess of deep  inelastic  scattering
events in  positron--proton  collisions  at high  $Q^2$ by both the H1
\cite{H1hiq}  and  ZEUS  \cite{ZEUShiq}  collaborations  at  HERA  has
prompted much  speculation  about  possible new physics  explanations.
Obvious  candidates  are  a  new  four--fermion   contact  interaction
$\Lambda^{-2} \bar e e \bar q q$ with $\Lambda \sim  \cO(1-2\tev)$, or
the production of a new heavy `leptoquark' resonance $e^+q \to \LQ \to
e^+q$ with $M_\LQ \sim 200\gev$ \cite{review}.  The electric charge of
such  an  object  is  not  yet  known,  but  if   $e_{\LQ}   =  +2/3$,
corresponding  to $e^+d \to \LQ$ for  example,  then the new  particle
could be a heavy  squark in an  $R$--parity  violating  supersymmetric
extension of the Standard Model.

It is important to  investigate  all possible  ways in which one could
distinguish between a conventional explanation (i.e.  a fluctuation of
the  standard  model  DIS  process)  and new  physics  scenarios.  One
diagnostic  tool which has already been  advocated  \cite{EKS}  in the
context of large $E_T$ jet  production  at hadron  colliders is to use
the    pattern    of    hadronic    energy    flow   in   the    event
[\ref{ref:DKT}--\ref{ref:LPHD}].   This   is   based   on   the   idea
\cite{DKT,DKMT}   that  the  overall  structure  of  particle  angular
distributions  in a hard scattering  process (the `event  portrait' or
`antenna  pattern') is governed by the underlying  colour  dynamics at
short  distance.  Thus  the  event  portrait  can  be  regarded  as  a
`partonometer' \cite{EKS} mapping the basic scattering process.

The events at high $x$ and $Q^2$ at HERA seem ideally suited to such a
study, being characterized by an energetic, well--separated lepton and
jet in the final state.  Furthermore the various candidate  underlying
$eq \to eq$ processes  ($t$--channel  $\gamma^*,Z$  exchange,  contact
interaction,   $s$--channel   colour--triplet   resonance  production,
\ldots)  have  distinctive  antenna  patterns,  as we  shall  see.  In
practice  one  could,  with  sufficiently   high  statistics,  use  an
additional  soft (gluon) jet as a probe of the antenna  pattern.  With
fewer events the  distribution  of soft  hadrons can be used  instead.
Both of these  quantities  are  related to the  inclusive  soft  gluon
distribution in the  next--to--leading  order $eq \to eqg$  processes,
the former  directly and the latter  through the  hypothesis  of Local
Parton  Hadron  Duality  (LPHD)   \cite{LPHD}  in  which  the  angular
distribution of soft particles emitted at wide angles to the energetic
jets follows that of the underlying  soft partons, with the rate being
determined  by  overall  multiplicative   energy--dependent  cascading
factors \cite{DKT,drag}.

The idea, then, is to use the angular  distributions of soft particles
or jets as a probe of new physics  contributions to high--$Q^2$ $e^+q$
scattering.  We  imagine  a  situation   where  a  larger   sample  of
(presumably standard model DIS) events at slightly lower $Q^2$ is used
as a control, to check the  approximate  validity of our  quantitative
predictions  for the antenna  pattern.  This can then be compared with
the observed  antenna  pattern for the sample of excess events.  As we
shall see, in some cases the `signal' and  `background'  distributions
can differ by factors  of 2 or more.  The  variation  of the  patterns
with the DIS variable $y$ will also be a useful discriminant.

We should also remark that experimental support for the feasibility of
such antenna  pattern  studies has come  recently  from the CDF and D0
collaborations  at the  Tevatron  $p \bar p$  collider  \cite{CDF,D0}.
Their analyses have shown that distinctive colour interference effects
in multijet and $W+$jet production survive the hadronization phase and
are clearly visible in the data.

In the following  section we derive the basic antenna  pattern results
for standard DIS and leptoquark production.  The case of a new contact
interaction  is  obtained  as  a  limiting  case  of  the  latter.  In
Section~3  we  present  numerical   predictions  for  various  typical
kinematic  configurations.  Analogous  results  for soft {\it  photon}
production  are obtained in  Section~4,  and  Section~5  contains  our
conclusions.

\section{Antenna patterns for deep inelastic scattering and
leptoquark production}

The distribution of soft gluon radiation is controlled by
the basic antenna pattern (see for example Ref.~\cite{book})
\beq
[ij] \equiv {p_i \cdot p_j \over p_i\cdot k\; p_j \cdot k}
 = { 1 - {\bf n}_i \cdot {\bf n}_j \over
\omega^2 ( 1 - {\bf n} \cdot {\bf n}_i)\;
( 1 - {\bf n} \cdot {\bf n}_j)} \; ,
\label{antenna}
\eeq
where the $p_i^\mu = E_i(1,{\bf n}_i)$
are the four--momenta of the energetic quarks and leptons
participating in the hard scattering process, and
$k^\mu=\omega(1,{\bf n})$ is the four--momentum of the soft
gluon. The radiation patterns presented below 
correspond to  the $\omega/E_i \to 0$ limits of the exact $eq \to eqg$
matrix elements.

We start by considering the Standard Model process $e(p_1) + q(p_2)
\to e(p_3) + q(p_4) $ by
$t$--channel $\gamma^*,Z$ exchange.
If the invariant mass of the $eq$ system is  $M$, and if the
angle between the incoming and outgoing quarks (in the $eq$ c.m.s.
frame) is $\Theta_q$, i.e. $\cos\Theta_q = {\bf n}_2\cdot {\bf n}_4$,
then the usual DIS variables are
\beq
x = \frac{M^2}{s}\; ,\quad y = \frac{1}{2}(1-\cos\Theta_q)\; , 
\quad Q^2 =
yM^2\; .
\label{DISvar}
\eeq
The scattering process with the various momenta labelled is shown in
Fig.~1.

Since our aim is to distinguish the patterns for resonance production
and the normal deep inelastic scattering, we  consider fixed $M$
and variable $\Theta_q$
($y$). For the Standard Model process the gluon energy and angular
distribution is simply (see for example \cite{book})
\beq
\frac{1}{\sigma_0} {d \sigma \over d\omega d\Omega_{\bf n}}
 = {\alpha_s C_F \over 4\pi^2} \;\omega\; \cF_{\rm SM}
\label{eq:basic}
\eeq
where
\beq
\cF_{\rm SM} = 2 [24] =
 {2 p_2 \cdot p_4 \over p_2\cdot k\; p_4 \cdot k}
 = {2( 1 - \cos\Theta_q) \over
\omega^2 ( 1 - \cos\theta_2)\;
( 1 - \cos\theta_4 ) } \; ,
\label{antenna24}
\eeq
where $\cos\theta_i = {\bf n}\cdot{\bf n}_i$ denotes the angle between
the soft gluon and the  corresponding  quark.  The gluon  emission  is
coherent, and depends on the relative  orientation of the incoming and
outgoing quark  directions.  Eq.~(\ref{antenna24})  can be interpreted
as a colour  `string'  connecting  the  incoming  and outgoing  quarks
\cite{drag,string}, and is closely related to the familiar result $\cF
= 2 [q\bar q]$ for the crossed  process $e^+e^- \to q \bar q$ (see for
example \cite{book}).

We now turn to the radiation pattern corresponding to the
production of an unstable colour--triplet, $s$--channel scalar resonance
 $\LQ$ of mass
$M$ and decay width $\Gamma$, i.e. $e q \to \LQ \to e q$.
We first note that the emission of a soft gluon off an on--shell
 colour--triplet
 {\it scalar boson} is described by the same factor as emission off
a colour--triplet {\it fermion}, i.e.
\beq
\cM^{(1)} \simeq \cM^{(0)} \; T^a_{ij} g_s
{ P\cdot
\epsilon_\lambda^a(k) \over P\cdot k}  \; ,
\eeq
where $T^a$ is a SU(3) colour  matrix,  $P^\mu$ is the momentum of the
emitting particle, and  $\epsilon^a_\lambda$ is the gluon polarization
vector.  We can therefore use results already obtained for heavy quark
production                          and                          decay
[\ref{ref:DKT},\ref{ref:book},\ref{ref:JIKIA}--\ref{ref:kohs}]      to
write down the result for leptoquark production and decay:\footnote{In
fact the soft gluon  distribution for $eq \to \LQ \to eq$ is identical
to that  for $W b \to t \to W b$ with  $m_t =  M_{\LQ}$,  $\Gamma_t  =
\Gamma_{\LQ}$ and $m_b = 0$.}
\beq
\cF_{\rm LQ} =  2 \left( [2P] + [4P]-  [PP]\right) +
2 \chi \left( [PP] + [24] - [2P] -[4P]  \right)
\label{antennalq}
\eeq
where $P = p_1 + p_2$ is the leptoquark momentum.
The factor $\chi$ in (\ref{antennalq}) is given by
\beq
\chi = {M^2\Gamma^2 \over (P\cdot k)^2 + M^2\Gamma^2}
= {\Gamma^2 \over \omega^2 + \Gamma^2}
\eeq
where the second expression corresponds to the $\LQ$ c.m.s. frame.
As discussed at length in Ref.~\cite{kos}, the radiation pattern depends,
through the factor $\chi$,
on the relative size of the gluon energy and the leptoquark decay width.
In this respect it is instructive to consider the two (formal)
limits $\Gamma\to \infty$ ($\chi \to 1$) and $\Gamma\to 0$
($\chi \to 0$), for fixed $\omega$.
In the former, the leptoquark decays immediately after
it is produced and has no time to radiate gluons of wavelength $\sim
1/\omega$. In this limit
\beq
\cF_{\rm LQ} \rightarrow 2 [24] \; ,
\label{antennalq1}
\eeq
which is identical to the standard DIS pattern (\ref{antenna24})
 corresponding to {\it coherent} emission.
In contrast, for $ \Gamma/\omega \to 0$ the emission takes place
on two very different timescales, corresponding to the
  {\it production} stage and the {\it decay} stage \cite{kos}:
\beq
\cF_{\rm LQ} \rightarrow \left\{ 2 [2P]- [PP]\right\} +
\left\{ 2 [4P] - [PP]\right\} \; .
\label{antennalq0}
\eeq

At threshold,  where there is essentially no radiation  from the heavy
leptoquark,  the  two  terms  in  $\{\}$   correspond  to  independent
radiation  off the  initial and final  state  (massless)  quarks,  see
(\ref{eq:indep})  below.  Note that it is  straightforward  to  verify
that the first term on the right--hand side of (\ref{antennalq0}) does
indeed correspond to the $k^\mu\to 0$ limit of the real gluon emission
matrix  element  squared  for  $e  + q  \to  \LQ  + g$  calculated  in
Refs.~\cite{ks,spira}.

With no {\it a priori}  knowledge  of the decay width of the new heavy
particle, the antenna pattern  (\ref{antennalq}) could in principle be
used to  obtain a  measurement.  This was the  approach  advocated  in
Ref.~\cite{kos}  for the top quark.  As we shall see in the  following
section, in certain regions of phase space the antenna pattern is very
sensitive to $\chi$, and therefore to $\Gamma$.  In practice, it seems
that for the class of  leptoquark  models  proposed  \cite{review}  to
explain the excess of  high--$Q^2$  events at HERA, the decay width is
rather  small.  In  particular,  a  scalar  leptoquark  coupling  with
strength  $\lambda$ to $eq$ has a corresponding  decay width $\Gamma =
M\lambda^2/(16\pi)$.  For  `first  generation'  leptoquarks  values of
$\lambda^2 <  \cO(10^{-2})$  are allowed by low--energy  data (see for
example  Ref.~\cite{ks}  and  references  therein).  This implies that
such  resonances  should be very narrow, i.e.  $\Gamma <  \cO(40\mev)$
for $M\sim  200\gev$.  If we are  interested in the  distributions  of
soft hadrons or jets with  energies of order a few GeV, then $\chi \ll
1$ and  (\ref{antennalq0})  is the  appropriate  distribution  for the
leptoquark signal.

Finally, we note that the antenna pattern for a
 $\bar e e \bar q q$ contact interaction corresponds to the
limit $\chi \to 1$, and is therefore identical to  the standard
DIS result, Eq.~(\ref{antenna24}).

\section{Numerical results}

In this section we present  numerical  results for the standard  model
DIS and  leptoquark  soft  gluon  distributions.  We work in the  $eq$
c.m.s.  frame  with  angles  defined  as in  Fig.~1,  and focus on the
dependence of the  dimensionless  quantity $\cN = \omega^2 \cF$, where
$\cF_{\rm SM}$ and $\cF_{\rm LQ}$ are defined in (\ref{antenna24}) and
(\ref{antennalq0})  respectively,  on the gluon  direction  ${\bf n}$.
Simple algebra gives
\bea
\label{eq:nsm}
\cN_{\rm SM} & = & {2(1-\cos\Theta_q) \over
(1-\cos\theta_2)(1-\cos\theta_4) } \; ,\\
\label{eq:indep}
\cN_{\rm LQ} & = & {1+\cos\theta_2 \over 1-\cos\theta_2}
+                  {1+\cos\theta_4 \over 1-\cos\theta_4} \; , \\
{ \cN_{\rm LQ}\over \cN_{\rm SM}}  & = & {1 -\cos\theta_2 \cos\theta_4
 \over 1-\cos\Theta_q} \; .
\label{eq:ratio}\eea
The  patterns  and their  ratio are  displayed  in  Figs.~2  and 3, as
functions of $\theta_g$  and  $\phi_g$, the polar and azimuthal  gluon
angles with  respect to the  incoming  quark  direction,\footnote{i.e.
$\theta_2         =         \theta_g,\          \cos\theta_4         =
\cos\phi_g\sin\theta_g\sin\Theta_q  + \cos\theta_g\cos\Theta_q$.}  and
for fixed values of $\Theta_q = 45^\circ,  90^\circ,  135^\circ$, i.e.
$ y=0.146,0.5,  0.854$.  To avoid the  collinear--singular  regions of
phase    space,    cuts    $\theta_2,\theta_4    >    10^\circ$    are
imposed.\footnote{The  cuts  on  $\theta_2,\theta_4$  are  omitted  in
Fig.~3,  since  the  ratios  are  finite  ($=1$) in the two  collinear
limits.}

We note the following points:
\begin{itemize}
\item[(i)]
For the standard model distribution, there is a significant enhancement
of radiation in the region between the
quark directions (i.e. $\phi \sim 0^\circ,\; 0^\circ \lapprox \theta
\lapprox \Theta_q$),  as expected.
This enhancement is largely absent in the $\LQ$ case, where the radiation
pattern is simply a superposition of independent radiation off the
initial and final state quarks.
\item[(ii)]
In the limit $\Theta_q \to 0^\circ$, $\cN_{\rm SM}$ vanishes everywhere
since the final state comoving colour triplet and antitriplet
behave as a colour singlet, whereas $\cN_{\rm LQ}$  is simply twice
the radiation off a single quark. In Ref.~\cite{dkos}, similar effects
where discussed for $e^+e^- \to t \bar t \to W^+W^- b \bar b$ production
at threshold.
\item[(iii)] For $\Theta_q = 90^\circ$ scattering, the ratio of the SM
and $\LQ$ distributions achieves its minimum and maximum values in the
plane of the scattering, thus $\cN_{\rm LQ} = \frac{1}{2} \cN_{\rm SM}
$    for     $(\phi_g,\theta_g)    =    (0^\circ,    45^\circ)$    and
$(180^\circ,135^\circ)$ and $\cN_{\rm LQ} = \frac{3}{2} \cN_{\rm SM} $
for     $(\phi_g,\theta_g)     =     (0^\circ,     135^\circ)$     and
$(180^\circ,45^\circ)$.  The  distributions  are the  same  for  gluon
directions in the planes perpendicular to ${\bf n}_2$ and ${\bf n}_4$,
i.e.  $\theta_2,\theta_4 = 90^\circ$.
\end{itemize}

Finally, from the above  discussion we would expect that the azimuthal
distribution  of soft  gluons  (hadrons)  around the final state quark
(jet) direction would be more uniform for quarks from leptoquark decay
than from standard deep inelastic scattering.  To see this, we show in
Fig.~4 the azimuthal  $\tilde\phi_g$  distribution of the gluon around
the final state quark direction ${\bf n}_4$, for $\Theta_q = 90^\circ$
and various fixed $\theta_4$.  A significant  azimuthal  asymmetry for
$\cN_{\rm  SM}$  is  observed  with  a  maximum  in the  plane  of the
scattering between the quark directions ($\tilde\phi_g = 0^\circ$), as
expected.  In  contrast,   the   dependence   of  $\cN_{\rm   LQ}$  on
$\tilde\phi_g$ is very weak, particularly for small $\theta_4$.

\section{Soft photon emission}

As discussed in the introduction, it would be of considerable interest
in distinguishing new physics models of the HERA high--$Q^2$ events
to know the electric charge of the quarks in the $eq\to eq$ process.
In principle, this information is contained in the distribution
of soft {\it photon} radiation, which can be obtained in an analogous
way to the soft gluon distributions of Section~2. The main difference 
is the presence of additional contributions from emission off the 
incoming and outgoing positrons.\footnote{The results in this section
are for $e^+ q\to e^+ q$ scattering. Those for $e^- q \to e^- q$
can be obtained by an appropriate change of sign.} The result is 
(cf. Eqs.~(\ref{eq:basic},\ref{antenna24},\ref{antennalq}))
\beq
\frac{1}{\sigma_0} {d \sigma \over d\omega_\gamma d\Omega_{\bf n}}
 = {\alpha \over 4\pi^2} \;\omega_\gamma\; \cF^\gamma
\eeq
where
\bea
\half \cF_{\rm SM}^\gamma &= & e_q^2 [24] - e_q \{ [12] + [34]
 - [14]  - [23] \} + [13] \label{eq:photsm}
\\
\half \cF_{\rm LQ}^\gamma &= & 
  e_q(1+e_q) \{[2P] +[4P]\}  + (1+e_q) \{ [1P] +[3P]\}
\nonumber \\ 
&&  - e_q \{ [12] + [34] \}  - (1+e_q)^2 [PP] 
\nonumber \\ 
&& + \chi \Bigl( (1+e_q)^2 [PP] - e_q(1+e_q) \{[2P]  +[4P]\} 
\nonumber \\ 
&& - (1+e_q) \{[1P]  +[3P]\} +  e_q^2 [24] + e_q \{[14]  +[23]\} + 
[13] \Bigr)
\label{eq:photlq}
\eea
and, as before, $\cF_{\rm SM}^\gamma = \cF_{\rm LQ}^\gamma(\chi=1)$.
As argued in the previous section, it is the $\chi \to 0$ limit
of $\cF_{\rm LQ}^\gamma$ which is relevant in practice, i.e. for photons
with energy $\omega_\gamma \gg \Gamma_{\LQ}$. In this limit we have
\bea
\half \cF_{\rm LQ}^\gamma &= & 
  e_q(1+e_q) \{[2P] +[4P]\}  + (1+e_q) \{ [1P] +[3P]\}
\nonumber \\ 
&&  - e_q \{ [12] + [34] \}  - (1+e_q)^2 [PP] 
\nonumber \\ 
&& =  H(\cos\theta_2) + H(\cos\theta_4)
\label{antennag0}
\eea
where
\beq
H(z) = {1 \over 1+z} + {e_q^2 \over 1-z} -\half(1+e_q)^2 \; .
\eeq
An interesting feature of the above distributions is the presence
of {\it radiation zeros} (see for example \cite{brodsky}),
 i.e. directions of the photon three--momentum
${\bf n}$ for which the cross section vanishes. To see this
for the distribution (\ref{antennalq}) we note that
\beq
H = 0 \qquad \mbox{for} \qquad z = z_0 \equiv {1-e_q \over 1+e_q} \; . 
\eeq
For the two cases of interest $e_q = \frac{2}{3}, -\frac{1}{3}$
for which $z_0 = \frac{1}{5}, 2$. Therefore only for $e^+u$ scattering is
the radiation zero in the physical region. For the full distribution
(\ref{antennag0}) to vanish we obviously require
\beq
 \cos\theta_2 = \cos\theta_4 = z_0 \; .
\eeq
Thus for $e^+u$  scattering  the  radiation  zero is in the  direction
given by the intersection of the two cones of half--angle  $\theta_0 =
\cos^{-1}(1/5)  = 78.46^\circ$  centred on the quark directions  ${\bf
n}_2$ and ${\bf n}_4$.  Three cases can be distinguished:
\begin{itemize}
\item[(i)] For $0^\circ < \Theta_q < 2 \theta_0$ there are two solutions,
corresponding to 
\beq
\theta_\gamma = \theta_0\; , \qquad 
\phi_\gamma = \pm \cos^{-1}\left({\tan(\Theta_q/2)
\over \tan\theta_0}\right) \; \label{eq:critical}.
\eeq
\item[(ii)] For $ \Theta_q = 2 \theta_0$ there is one solution,
\beq
\theta_\gamma = \theta_0\; , \qquad \phi_\gamma = 0^\circ \; ,
\eeq
corresponding to the bisector of the quark directions in the scattering plane.
\item[(iii)] For $ \Theta_q =0^\circ$ there is a cone of solutions
corresponding to $\theta_\gamma = \theta_0$. 
\end{itemize}

Although the above results on the location of the radiation zeroes
 have been derived for the leptoquark
radiation pattern, they apply equally well for the standard model
distribution (\ref{eq:photsm}), or indeed for the generic distribution
(\ref{eq:photlq}) for arbitrary $\chi$. This follows from the fact that
the zeroes are the result of completely destructive interference between
the classical electric fields associated with the different charged
particles, see for example Ref.~\cite{DKMT}.
They depend only on the relative orientation of the various
particles, irrespective of whether intermediate resonances are formed.
The key point to note is that since presumably the leptoquark
couples to either $u$ or $d$ but not both, the  radiation zero
is either fully present or completely absent.   In contrast, the
standard model background is a linear combination (determined by the
parton distributions) of $u$-- and $d$--type distributions, giving
a dip rather than a zero.

As a numerical illustration of these results, we show in Fig.~5 the 
antenna patterns $\cN^\gamma_{\rm SM}$, $\cN^\gamma_{\rm LQ}(e_q = 2/3)$
and $\cN^\gamma_{\rm LQ}(e_q = -1/3)$, with $\Theta_q = 90^\circ$.
To exhibit the radiation zeroes more clearly, Fig.~6 shows
the $\phi_\gamma$ dependence of the leptoquark $e_q = 2/3, -1/3$
distributions at the critical polar angle $\theta_\gamma = \theta_0$,
i.e. the slices through the two--dimensional distributions of Fig.~5
at this value of $\theta_\gamma$.  The two zeroes of the $e^+u$
distribution at the $\phi_\gamma$ angles given by Eq.~(\ref{eq:critical})
are clearly visible. Note also that the behaviour of the distributions
near the positron and the quark jet directions simply reflects the
magnitude of the charge of the corresponding particles.

\section{Conclusions}

If  the  observation  of an  excess  of  high--$Q^2$  events  at  HERA
persists, it will be important to devise new analysis  techniques  for
identifying  the origin of the  excess.  In this  paper we have  shown
that the angular  distribution of the accompanying  hadronic radiation
--  the  antenna  pattern  --  is  a  potentially  powerful  tool  for
discriminating  standard deep inelastic  scattering  events from those
arising  from  the  production  of  a  long--lived   coloured   scalar
`leptoquark'   resonance.  The  main  qualitative  difference  is  the
absence for the latter of an enhancement of hadronic radiation between
the incoming and outgoing  quark jet  directions  (string  effect), as
shown in Fig.~2.  It follows that soft  hadrons are  distributed  more
uniformly  in azimuth  around the final state quark jet  direction  in
events where a leptoquark  is produced, see Fig.~3.  Our  quantitative
predictions are based on the  phenomenologically  successful principle
of Local Parton Hadron  Duality, and should  therefore be a good guide
to the behaviour of the  distributions of soft hadrons and jets in the
detectors.  Ultimately,  however,  there  will  be no  substitute  for
detailed  Monte Carlo  studies  based on  parton--shower/hadronization
models,  provided  that these  include the correct  underlying  colour
structure.

Finally we have extended our results to include soft photon radiation.
Here the distributions have an additional  sensitivity to the electric
charge  of  the   leptoquark,   which  is  a  crucial   parameter   in
distinguishing  models.  For the  case of  charge  $5/3$  leptoquarks,
produced   for  example  in  $e^+u$   collisions,   the  soft   photon
distribution  contains  radiation zeroes.  These are absent for charge
$2/3$ leptoquarks produced in $e^+d$ collisions.

\vspace{1.0cm}
\noindent {\bf Acknowledgements} \\

We  thank  John  Ellis  and  Valery  Khoze  for  useful  comments  and
discussions.  This work was  supported in part by the UK PPARC and the
EU Programme ``Human Capital and Mobility'', Network ``Physics at High
Energy   Colliders'',   contract   CHRX-CT93-0357  (DG  12  COMA).  MH
gratefully   acknowledges   financial   support   in  the  form  of  a
DAAD-Dok\-tor\-an\-den\-sti\-pen\-di\-um (HSP III).

%%%%%%%%%%%%%%%%%%%%%%%%%%%%%%%%% FIGURE 1

\begin{figure}[t]
\begin{center}
\mbox{\epsfig{figure=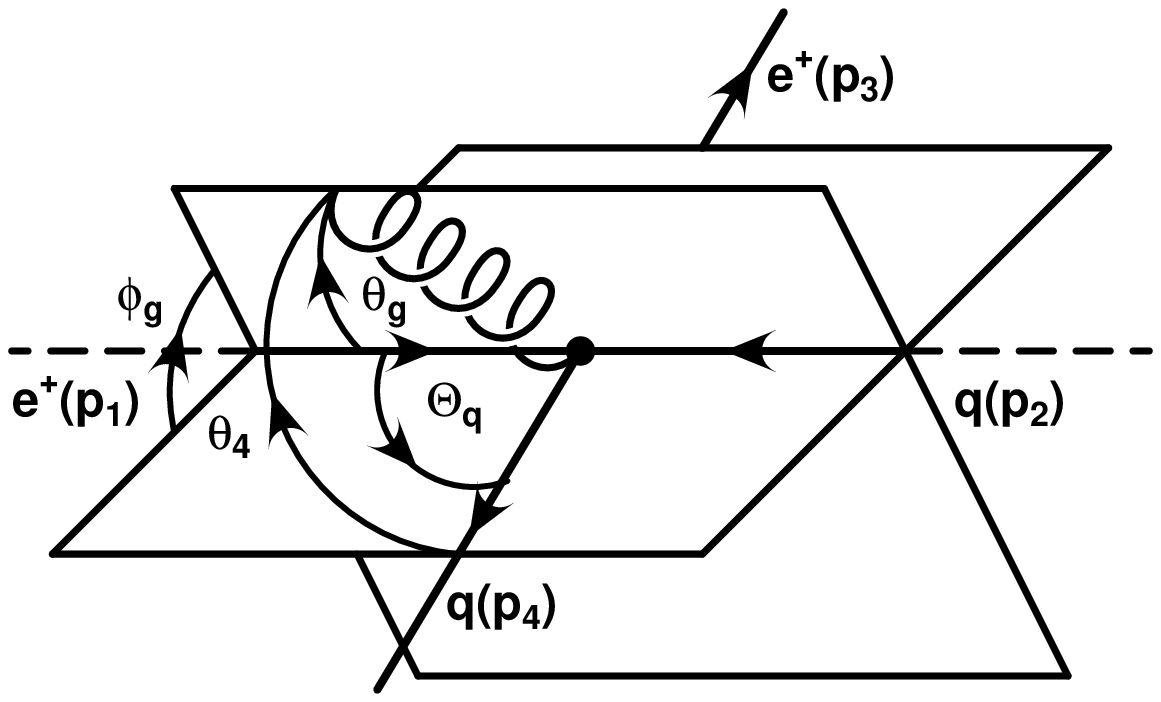,width=12cm}}
\caption[]{Parametrization of the kinematics for $e^+(p_1)
q(p_2) \rightarrow e^+(p_3) q(p_4) + g(k)$ scattering in the $e^+q$
c.m.s. frame. The orientation of the soft gluon relative
to the scattering plane is denoted by $\theta_g$ and
$\phi_g$ or, alternatively,
by the  angles with respect to  the directions of the participating
quarks: $\theta_4$ and $\theta_2 = \theta_g$. }
\label{fig.1}
\end{center}
\end{figure}

%%%%%%%%%%%%%%%%%%%%%%%%%%%%%%%%% FIGURE 2

\begin{figure}[t]
\begin{center}
\mbox{\epsfig{figure=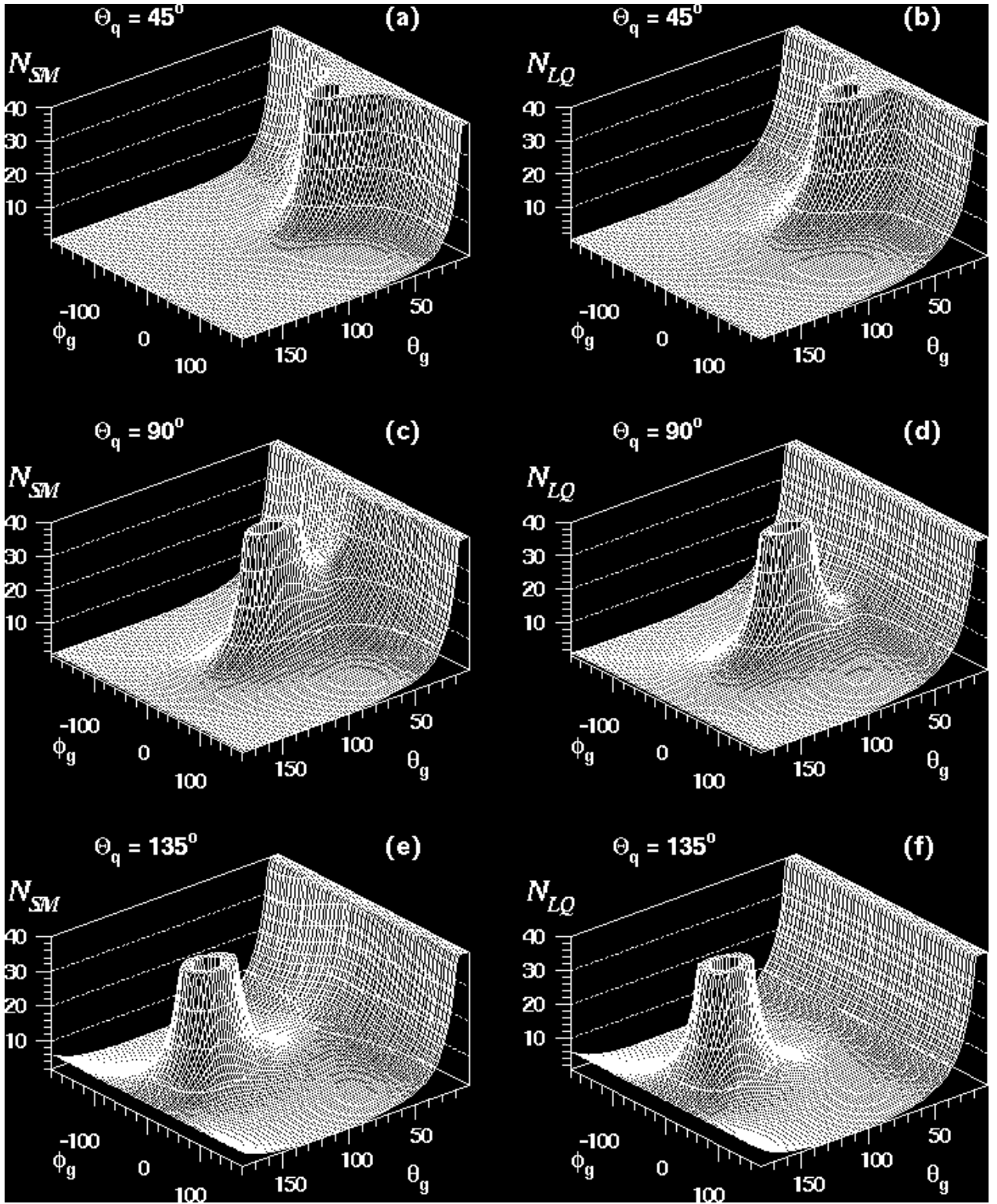,width=17.0cm}}
\caption[]{The dimensionless antenna patterns ${\cal N}_{\rm SM} = \omega^2
{\cal F}_{\rm SM}$ [(a),(c),(e)] and
${\cal N}_{\rm LQ} = \omega^2{\cal F}_{\rm LQ}$ [(b),(d),(f)]
of  Eqs.~(\ref{eq:nsm},\ref{eq:indep}) for different c.m.s. scattering
angles $\Theta_q$ (cf. Fig.~\ref{fig.1}). Note the cut of $10^\circ$
imposed around the incoming and outgoing quark directions.}
\label{fig.2}
\end{center}
\end{figure}

%%%%%%%%%%%%%%%%%%%%%%%%%%%%%%%%% FIGURE 3

\begin{figure}[t]
\begin{center}
\mbox{\epsfig{figure=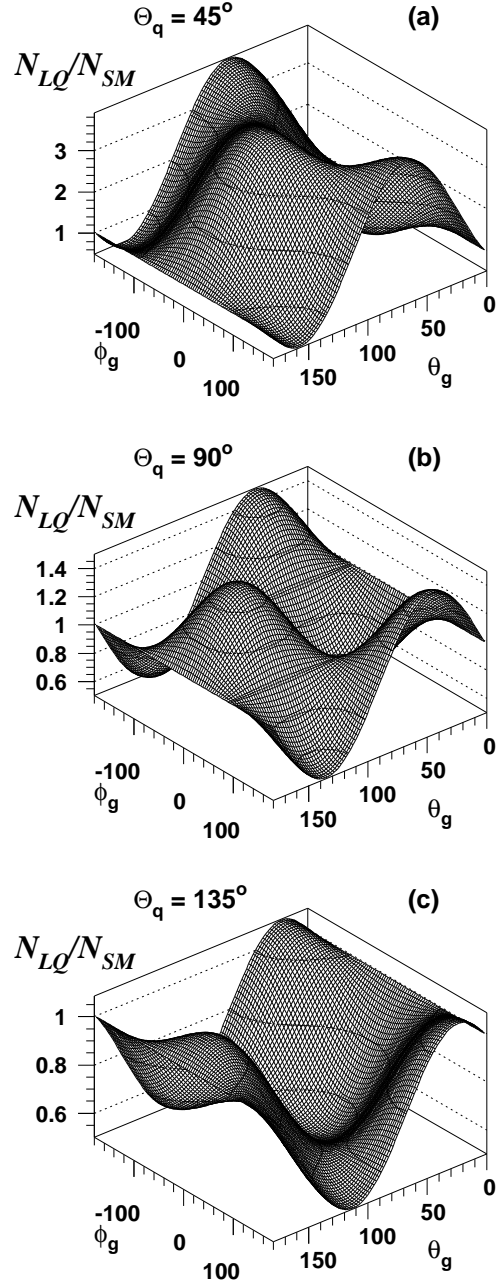,width=8.1cm}}
\caption[]{The ratios ${\cal N}_{\rm LQ}/{\cal N}_{\rm SM}$ of
the distributions in
Fig.~\ref{fig.2} for the three different c.m.s. scattering
 angles. In this case no angular cuts have been imposed.}
\label{fig.3}
\end{center}
\end{figure}

%%%%%%%%%%%%%%%%%%%%%%%%%%%%%%%%% FIGURE 4

\begin{figure}[t]
\begin{center}
\mbox{\epsfig{figure=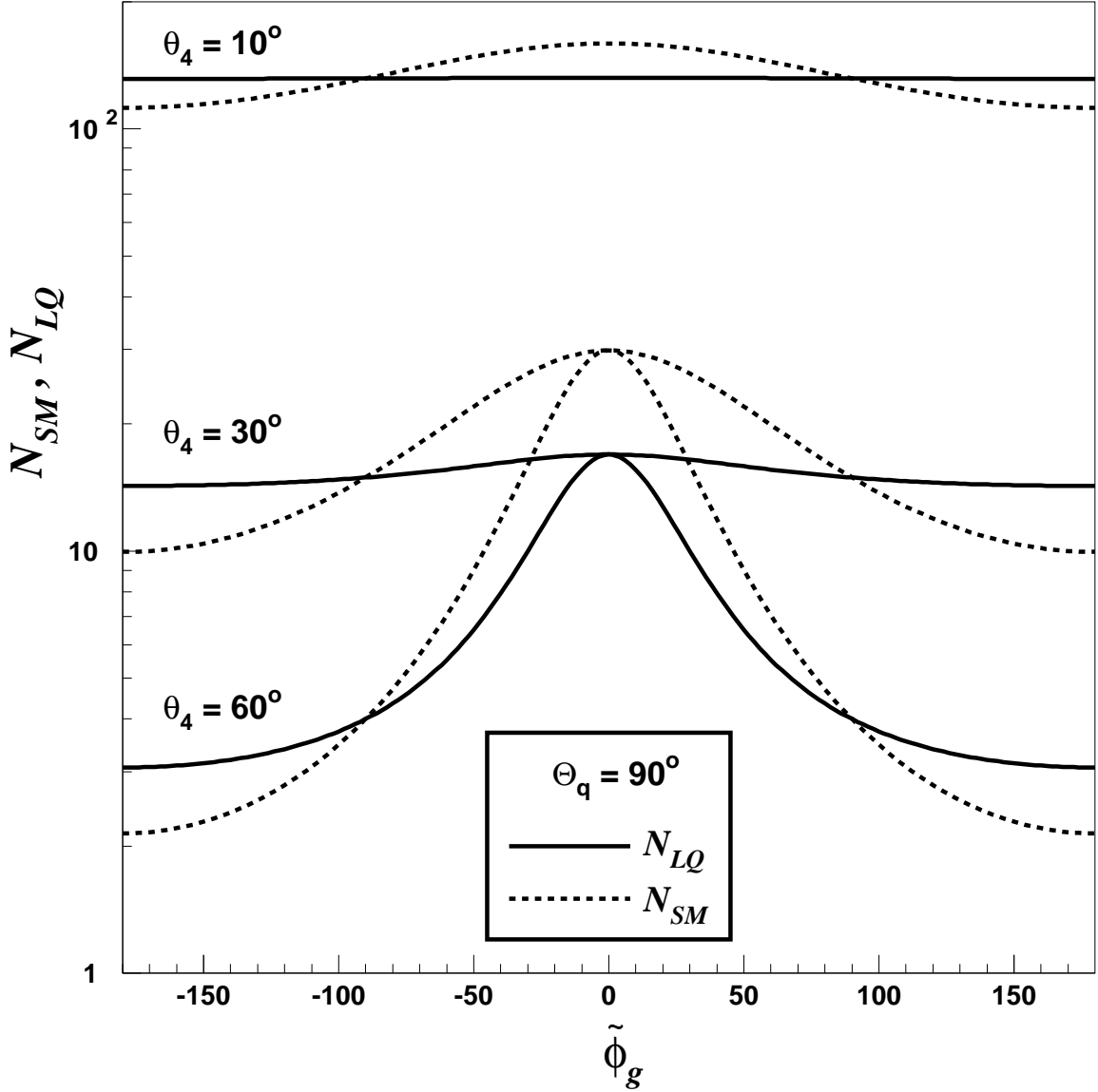,width=17.0cm}}
\caption[]{The dependence of the  antenna patterns
${\cal N}_{\rm SM}$ and ${\cal N}_{\rm LQ}$ on the azimuthal angle
$\tilde{\phi}_g$ of the soft gluon around the outgoing quark
$q(p_4)$. The gluon direction describes a cone around
the quark of half--angle $\theta_4$.
The direction of the incoming quark $q(p_2)$ is
defined by  $\tilde{\phi}_g = 0^\circ$, and
 the incoming positron $e^+(p_1)$ is at $\tilde{\phi}_g = \pm 180^\circ$.
The overall c.m.s. scattering
angle is fixed at $\Theta_q=90^\circ$.}
\label{fig.4}
\end{center}
\end{figure}

%%%%%%%%%%%%%%%%%%%%%%%%%%%%%%%%% FIGURE 5

\begin{figure}[t]
\begin{center}
\mbox{\epsfig{figure=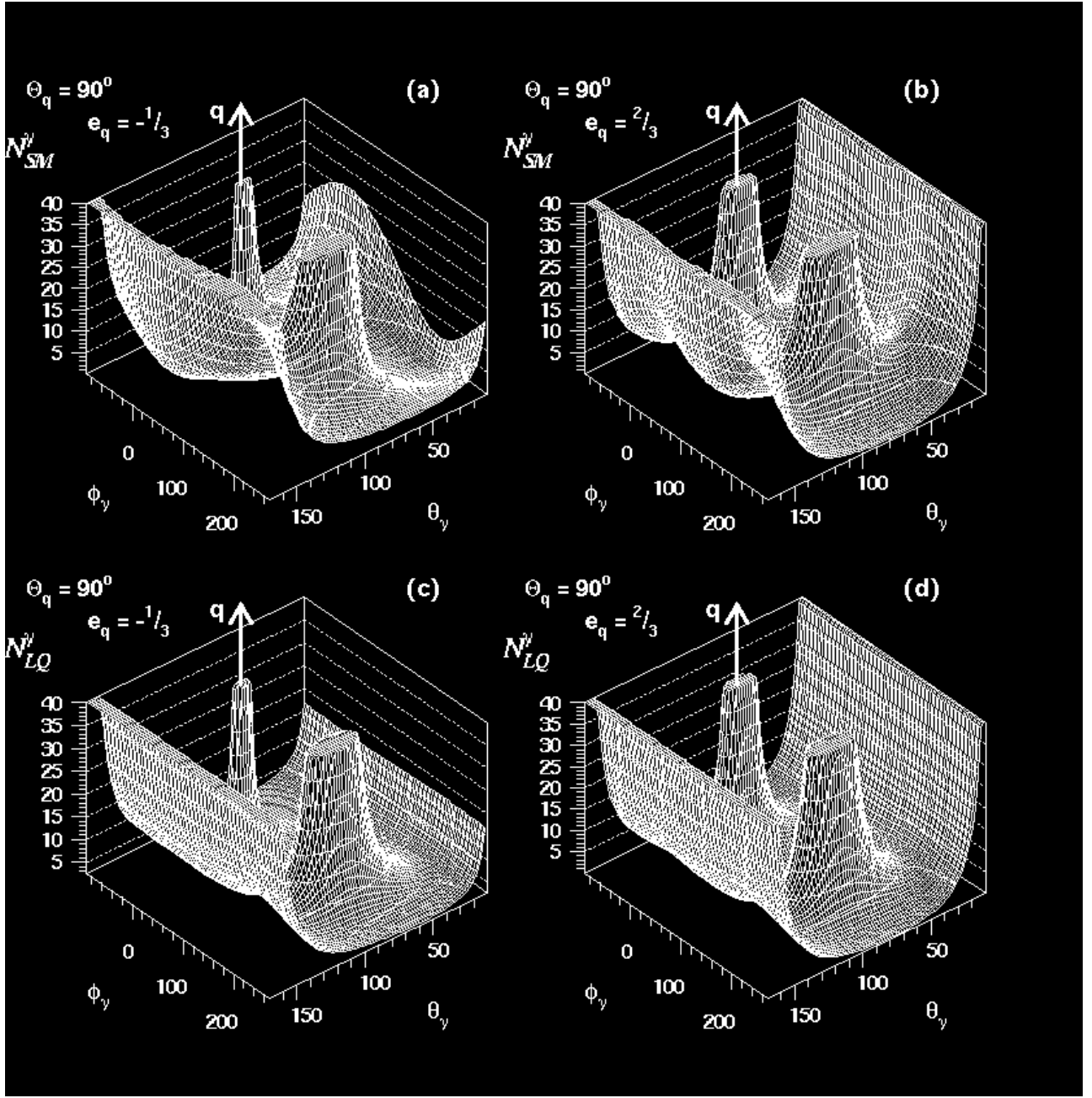,width=17.0cm}}
\caption[]{The pattern of  soft $\gamma$ radiation according to
Eqs.~(\ref{eq:photsm},\ref{antennag0}) with ${\cal N}^\gamma_{\rm SM,LQ} 
= \omega_\gamma^2
{\cal F}^\gamma_{\rm SM,LQ}$, for
$e^+d$ scattering [(a),(c)] and  $e^+u$ scattering [(b),(d)].
The overall c.m.s. scattering
angle is fixed at $\Theta_q=90^\circ$.}
\label{fig.5}
\end{center}
\end{figure}

%%%%%%%%%%%%%%%%%%%%%%%%%%%%%%%%% FIGURE 6

\begin{figure}[t]
\begin{center}
\mbox{\epsfig{figure=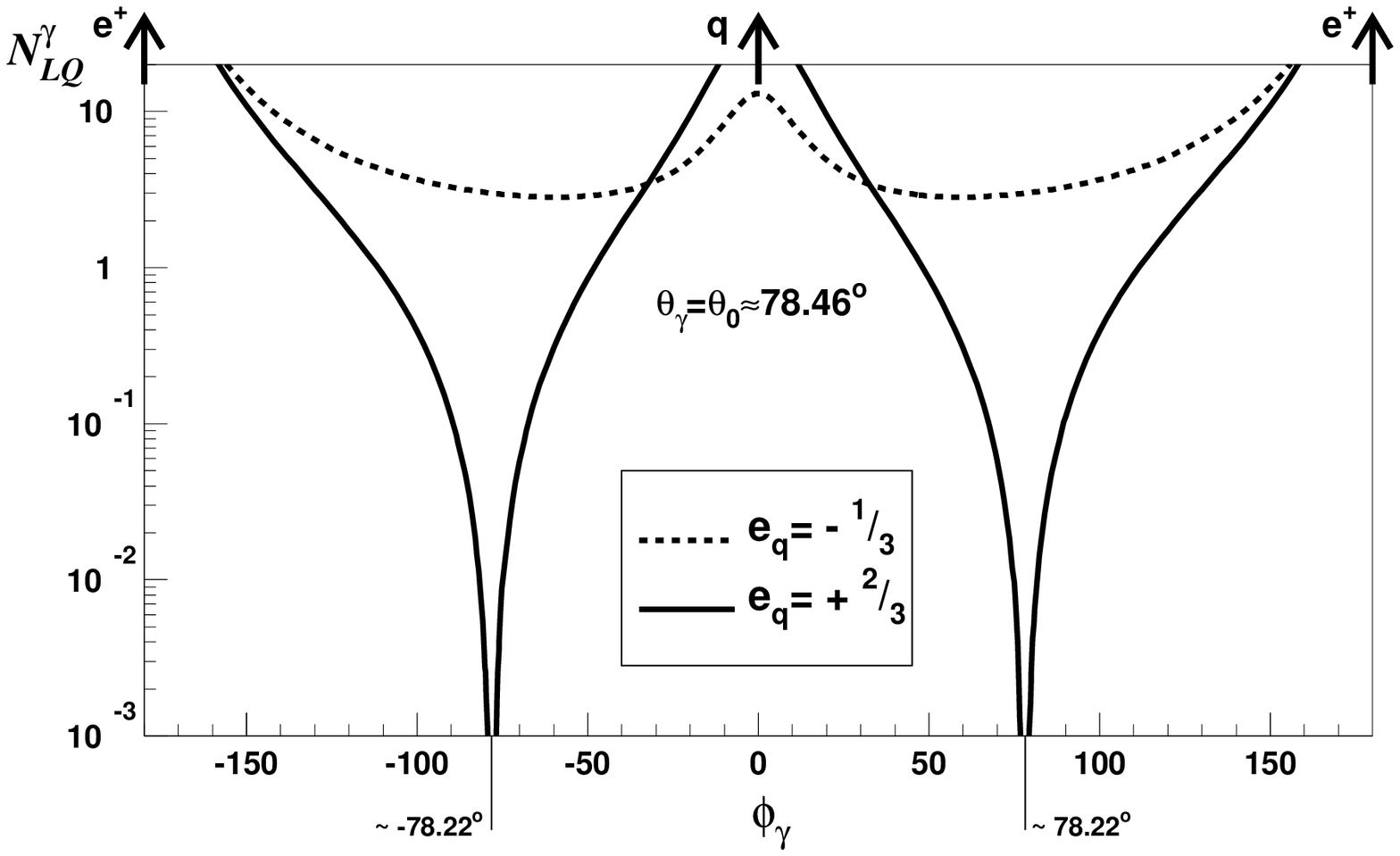,width=17.0cm}}
\caption[]{The soft photon antenna pattern ${\cal N}^\gamma_{\rm LQ}$ 
for $\Theta_q=90^\circ$ at the
 critical angle $\theta_\gamma = \theta_0 \simeq 78.46^\circ$
for $e^+u$ (solid line) and
$e^+d$ (dashed line) scattering. Note the radiation zeros at
$\phi_\gamma\simeq 78.22^\circ$ (cf. Eq.~(\ref{eq:critical})). The positions
of the $e^+$ and $q$ jets are indicated. Note also that 
${\cal N}^\gamma_{\rm SM}$ shows quantitatively the same behaviour
for this choice of $\theta_\gamma$.}
\label{fig.6}
\end{center}
\end{figure}


\newpage
\begin{thebibliography}{99}

\bibitem{H1hiq}\label{ref:H1hiq} 
H1 Collaboration: C.~Adloff et al.,  DESY preprint 97--24,
hep-ex/9702012 (1997).

\bibitem{ZEUShiq}\label{ref:ZEUShiq} 
ZEUS Collaboration: J.~Breitweg et al., DESY preprint 97--25,
hep-ex/9702015  (1997).

\bibitem{review}\label{ref:review}For a general discussion of the various
new physics possibilities see for example: \\
G.~Altarelli, J.~Ellis, G.F.~Giudice, 
S.~Lola and M.L.~Mangano,  preprint CERN--TH/97--40,
hep-ph/9703276  (1997).\\
J.L.~Hewett and T.G.~Rizzo, SLAC preprint SLAC-PUB-7430, 
hep-ph/9703337 (1997).

\bibitem{EKS}\label{ref:EKS} 
J.~Ellis, V.A.~Khoze and W.J.~Stirling, 
CERN preprint CERN--TH/96--225, hep-ph/9608486 (1996)
to be published in Zeit. Phys. 

\bibitem{DKT}\label{ref:DKT}  Yu.L.~Dokshitzer, V.A.~Khoze and
S.I.~Troyan, in Proc. 6th Int. Conf. on Physics in Collision,
ed. M.~Derrick (World Scientific, Singapore, 1987), p.417. \\
Yu.L.~Dokshitzer, V.A.~Khoze and
S.I.~Troyan, Sov. J. Nucl. Phys. {\bf 46} (1987) 712.
   
\bibitem{DKMT}\label{ref:DKMT} 
Yu.L.~Dokshitzer, V.A.~Khoze,
A.H.~Mueller and S.I.~Troyan, Rev. Mod. Phys. {\bf 60} (1988) 373.\\
Yu.L.~Dokshitzer, V.A.~Khoze and
S.I.~Troyan in:  Advanced Series on Directions in High Energy
Physics, Perturbative Quantum Chromodynamics, ed.
A.H.~Mueller (World Scientific, Singapore), v. 5 (1989) 241.

\bibitem{book}\label{ref:book} 
Yu.L.~Dokshitzer, V.A.~Khoze,
A.H.~Mueller and S.I.~Troyan, \lq\lq Basics of Perturbative
QCD", ed. J.~Tran Thanh Van, Editions Fronti\'{e}res,
Gif-sur-Yvette, 1991.

\bibitem{emw}\label{ref:emw}
 R.K.~Ellis, G.~Marchesini and
B.R.~Webber, Nucl. Phys. {\bf B286} (1987) 643; Erratum
 Nucl. Phys. {\bf B294} (1987) 1180.\\
R.K.~Ellis, presented at ``Les Rencontres de
Physique de la Vallee d'Aoste'', La Thuile, Italy, March 1987, preprint
FERMILAB--Conf--87/108--T (1987). 

\bibitem{DKTSJNP}\label{ref:DKTSJNP} 
Yu.L.~Dokshitzer, V.A.~Khoze and
S.I.~Troyan, Sov. J. Nucl. Phys. {\bf 50} (1989) 505.

\bibitem{DKS}\label{ref:DKS} 
Yu.L.~Dokshitzer, V.A.~Khoze and T.~Sj\"{o}strand,
 Phys. Lett. {\bf B274} (1992) 116.

\bibitem{MarWeb}\label{ref:MarWeb}
G.~Marchesini and B.R.~Webber, Nucl. Phys. {\bf B330} (1990) 261.

\bibitem{ZEPPEN}\label{ref:ZEPPEN} 
D.~Zeppenfeld, Madison preprint MADPH--95--933 (1996).

\bibitem{KSW}\label{ref:KSW}
V.A.~Khoze and W.J.~Stirling, Durham preprint DTP/96/106,
hep-ph/9612351 (1996). 

\bibitem{LPHD}\label{ref:LPHD}
Ya.I.~Azimov, Yu.L.~Dokshitzer, 
V.A.~Khoze and S.I.~Troyan, Z. Phys. {\bf C27} (1985) 65;
{\bf C31} (1986) 213.

\bibitem{drag}\label{ref:drag} 
Ya.I.~Azimov, Yu.L.~Dokshitzer,
V.A.~Khoze and S.I.~Troyan, Phys. Lett. {\bf B165} (1985) 147;
Sov. J. Nucl. Phys. {\bf 43} (1986) 95.

\bibitem{CDF}\label{ref:CDF} 
CDF Collaboration:  F.~Abe et al., Phys. Rev. {\bf D50} (1994) 5562;
P.~Giannetti, FERMILAB preprint FERMILAB-CONF-94-151-E, presented
at the 27th International Conference on 
High Energy Physics (ICHEP), Glasgow, 
Scotland, July 1994.
 
\bibitem{D0}\label{D0} 
D0 Collaboration: D.E.~Cullen-Vidal, presented at the
Annual Divisional Meeting ({\it DPF96}) of the APS Division of 
Particles and  Fields, Minneapolis, August 1996,  FERMILAB
preprint, FERMILAB Conf--96--304--EI (1996);
H.~Melanson, presented
at INFN Eloisatron project 33rd Workshop, Erice, Italy, October 1996.

\bibitem{string}\label{ref:string} 
B.~Andersson, G.~Gustafson and
T.~Sj\"{o}strand, Phys. Lett. {\bf B94} (1980) 211.

\bibitem{JIKIA}\label{ref:JIKIA}
G.~Jikia, Phys. Lett. {\bf B257} (1991) 196.

\bibitem{DKTlun92}\label{ref:DKTlun92} 
Yu.L.~Dokshitzer, V.A.~Khoze and S.I.~Troyan, University of Lund
preprint LU--TP--92--10 (1992).

\bibitem{kos}\label{ref:kos}
V.A.~Khoze, L.H.~Orr  and W.J.~Stirling, Nucl. Phys. {\bf B378} (1992) 413.

\bibitem{dkos}\label{ref:dkos}
Yu.L.~Dokshitzer, V.A.~Khoze, L.H.~Orr  and W.J.~Stirling,
Nucl. Phys. {\bf B403} (1993) 65.

\bibitem{kohs}\label{ref:kohs}
V.A.~Khoze,  J.~Ohnemus and W.J.~Stirling,
Phys. Rev. {\bf D49} (1994) 1237. 


\bibitem{ks}\label{ref:ks}
Z.~Kunszt and W.J.~Stirling, preprint DTP/97/16, hep-ph/9703427 (1997).

\bibitem{spira}\label{ref:spira}
T.~Plehn,   H.~Spiesberger, M.~Spira and  P.M.~Zerwas, 
preprint DESY-97-43, hep-ph/9703433 (1997).



\bibitem{brodsky}\label{ref:brodsky}
S.J.~Brodsky  and R.W.~Brown,  Phys.  Rev. Lett. {\bf 49} (1982) 966.


\end{thebibliography}
\end{document}